\documentclass[aps,prd,superscriptaddress,twocolumn,secnumarabic,nofootinbib]{revtex4-1}
\pdfoutput=1
\usepackage[utf8]{inputenc}
\usepackage[normalem]{ulem} 
\usepackage[english]{babel}%,frenchb
\usepackage{tabularx}
\usepackage{array}
\usepackage{graphics}
\usepackage[pdftex]{graphicx}
\usepackage{mathtools}
\usepackage{amssymb}
\usepackage{amsmath}
\usepackage{slashed}
\usepackage{braket}
\usepackage[usenames,dvipsnames]{xcolor}

\usepackage{float}

\usepackage{bm}

\usepackage{verbatim}

\usepackage{hyperref}
\hypersetup{colorlinks,citecolor=nicegreen,linkcolor=niceblue}
\hypersetup{colorlinks=true}

\newcommand{\cb}{{\mathcal B}}
\newcommand{\bea}{\begin{eqnarray}}
\newcommand{\eea}{\end{eqnarray}}
\newcommand{\beq}{\begin{equation}}
{
\newcommand{\eeq}{\end{equation}}
\newcommand{\ec}{\end{center}}
\newcommand{\bc}{\begin{center}}

\newcommand{\gev}{{\rm GeV}}
\newcommand{\mev}{{\rm MeV}}

\newcommand{\pdir}{p\kern -5.2pt\raise 0.2ex\hbox {/}}

\newcommand{\vdir}{v\kern -5.75pt\raise 0.15ex\hbox {/}}
\newcommand{\kdir}{k\kern -5.75pt\raise 0.15ex\hbox {/}}
\newcommand{\epsdir}{\epsilon\kern -5.0pt\raise 0.15ex\hbox {/}}
\newcommand{\bvdir}{\bar{v}\kern -5.75pt\raise 0.15ex\hbox {/}}
\newcommand{\Ddir}{D\kern -7.75pt\raise 0.20ex\hbox {/}}
\newcommand{\Adir}{A\kern -7.75pt\raise 0.20ex\hbox {/}}
\newcommand{\ldir}{l\kern -5.0pt\raise 0.2ex\hbox{/}}
\newcommand{\varepsdir}{\varepsilon\kern -5.5pt\raise 0.15ex\hbox{/}}

\newcommand{\nn}{\nonumber}

\newcommand {\E}[1]{\times 10^{#1}}	% exponent notation
\newcommand {\e}[1]{\mathrm{\,#1}}       % units
\newcommand{\mc}[1]{\mathcal{#1}}
\newcommand{\mrm}[1]{\mathrm{#1}}

\newcommand{\re}[0]{\mrm{Re}}
\newcommand{\im}[0]{\mrm{Im}}
\newcommand{\cres}[0]{C_9^\mrm{res}(q^2)}
\newcommand{\delj}[0]{\delta_j}
\newcommand{\acp}[0]{\mathcal{A}_\mathrm{CP}}
\makeatother

\definecolor{niceblue}{rgb}{0.15,0.15,0.6}
\definecolor{nicegreen}{rgb}{0.1,0.5,0.1}
\definecolor{Red}{rgb}{1.,0.,0.}

\definecolor{Green}{rgb}{0.2,.7,0.2}

\begin{document}
\title{Enhanced CP asymmetries in $B \to K \mu^+ \mu^-$}

\author{Damir Be\v cirevi\' c} \email[Electronic address:]{\ damir.becirevic@ijclab.in2p3.fr}
\affiliation{IJCLab, P\^ole Th\'eorie (B\^at.~210), CNRS/IN2P3 et Universit\'e Paris-Saclay, 91405 Orsay, France}

\author{Svjetlana Fajfer} \email[Electronic address:]{\ svjetlana.fajfer@ijs.si}
\affiliation{Department of Physics, University of Ljubljana, Jadranska 19, 1000 Ljubljana, Slovenia}
\affiliation{Jo\v zef Stefan Institute, Jamova 39, P.\ O.\ Box 3000, 1001
  Ljubljana, Slovenia}

\author{Nejc Ko\v snik} \email[Electronic address:]{\ nejc.kosnik@ijs.si}
\affiliation{Department of Physics, University of Ljubljana, Jadranska 19, 1000 Ljubljana, Slovenia}
\affiliation{Jo\v zef Stefan Institute, Jamova 39, P.\ O.\ Box 3000, 1001
  Ljubljana, Slovenia}

\author{Aleks Smolkovi\v c} \email[Electronic address:]{\ aleks.smolkovic@ijs.si}
\affiliation{Jo\v zef Stefan Institute, Jamova 39, P.\ O.\ Box 3000, 1001
  Ljubljana, Slovenia}

%%%%%%%%%%%%%%%%%%%%%%%%%%%%%%%%%%%%%%%%%%%%%%%%%%
\begin{abstract}
We show that the current values of $R_K^\mathrm{exp}$ and $R_{K^\ast}^\mathrm{exp}$ can be accommodated by allowing a nonzero New Physics coupling $\delta C_9^{\mu\mu}$ to be complex, both in the scenario in which only $\delta C_9^{\mu\mu}$ is affected, and in the scenario with complex $\delta C_{9,10}^{\mu\mu}$ satisfying $\delta C_{9}^{\mu\mu}=-\delta C_{10}^{\mu\mu}$. A presence of the weak CP-violating phase can then be tested by measuring the CP-asymmetry, $\acp$.  We show that this asymmetry is enhanced around the peak of each $c\bar c$-resonance, and in fact more pronounced in the close vicinity of $J/\psi$ and $\psi(2S)$.  Therefore, measuring $\acp$ before and after the resonances' peak could be revelatory of the CP-violation that originates from beyond the Standard Model, or to be a significant constrain when building a realistic scenario of New Physics. 
\end{abstract}
%%%%%%%%%%%%%%%%%%%%%%%%%%%%%%%%%%%%%%%%%%%%%%%%%%

\maketitle

\section{Introduction}

One of the most intriguing observations made at LHCb so far is the indication of the lepton flavor universality violation (LFUV). 
Regarding the $b\to s\ell\ell$ processes it was found that the measured $R_{K^{(\ast)}} = {\cb' (B\to K^{(\ast)} \mu\mu)}/{\cb' ( B\to
  K^{(\ast)} e e)}$, where $\cb'$ stands for the partial branching fraction, is smaller than predicted in the Standard Model (SM). 
For example, in the bin of $q^2 \in [1.1,6]\e{GeV}^2$ the measured values, 
$R_K = 0.846^{+0.062}_{-0.056}$~\cite{Aaij:2014ora,Aaij:2019wad} and $R_{K^\ast} = 0.69^{+0.12}_{-0.09}$~\cite{Aaij:2017vbb}, are both about $2.5\,\sigma$ smaller 
than their SM estimate, $R_{K^{(\ast)}}=1.00(1)$~\cite{Bordone:2016gaq}.

To understand the origin of such a discrepancy between theory and experiment one readily extends the effective field theory beyond the SM 
and from a fit to the experimental data extracted from the full angular distribution of $B\to K^{(\ast)} \mu\mu$~\cite{Aaij:2015oid}, 
one can deduce which scenario of New Physics is preferred~\cite{Aebischer:2018iyb,Arbey:2019duh,Alguero:2019ptt}.
On the basis of such analyses and the available experimental data it is reasonable to assume that the LFUV in $R_{K^{(\ast)}}^\mathrm{exp}<R_{K^{(\ast)}}^\mathrm{SM}$ 
is due to a more pronounced coupling of New Physics to the muon pair in the final state. Being much smaller, such a coupling to the electron pair can be assumed to be zero. 
Of all the possibilities tested through the global analyses of the exclusive $b\to s\ell\ell$ modes the most 
favored sce\-na\-rios are those in which the signs of physics beyond the Standard Model (BSM) arise from coupling to the vector muonic current, i.e. $\delta C_9\neq 0$, and the one in which $ \delta C_9=-\delta C_{10}\neq 0$, where $\delta C_i\equiv  \delta C_i^{\mu\mu} = C_i^{\mu\mu}  - C_i^\mathrm{SM}$ stands for the contribution arising solely from BSM. 
In order to keep the number of free parameters minimalistic a common assumption is that $\delta C_i \in \mathbb{R}$. Starting from results based on the effective field theory approach 
one can then build a specific model verifying either $\delta C_9 \neq 0$ or $ \delta C_9 =-\delta C_{10} \neq 0$.

Of course, the assumption $\delta C_i \in \mathbb{R}$ helps keeping the number of BSM (real) parameters minimalistic but that assumption should be scrutinized experimentally. A physical consequence of allowing $\delta C_i \in \mathbb{C}$ is that the BSM effects of CP violation can be visible from the di-muon invariant mass spectra of the decays we consider here. A presence of the imaginary part in the BSM contribution to the Wilson coefficient(s) could be tested through various CP-asymmetries which are known to be tiny in the SM~\cite{Bobeth:2008ij,Bobeth:2011gi,Alok:2011gv,Becirevic:2012fy,Alok:2017jgr,Fleischer:2017yox}. LHCb already attempted measuring direct CP asymmetry
\bea\label{Acp}
\acp^{K^{(\ast )}} = {\cb (\overline B \to  \overline K^{(\ast )}  \mu\mu)-  \cb ( B \to   K^{(\ast )}  \mu\mu) \over \cb (\overline B \to  \overline K^{(\ast )}  \mu\mu)+  \cb ( B \to   K^{(\ast )}  \mu\mu)} \,,
\eea
and reported~\cite{Aaij:2014bsa}:~\footnote{Throughout this paper we will use $\acp$ both to denote the function $\acp(q^2)$ and its value obtained after appropriate integration in $q^2$, such as the value given in Eq.~\eqref{eq:acpexp}. This should not be a source of confusion as the distinction will be evident in the discussion.}
\bea\label{eq:acpexp}
\acp^{K^+} =0.012(17)(1),\quad \acp^{K^{0 \ast }} =-0.035(24)\,,
\eea
which suggest that the imaginary part of the Wilson coefficient(s) is likely to be small and is currently consistent with zero. 
Measuring this quantity is complicated either because it is small, or because it varies within a bin so that its integrated value over the size of one bin is very small. Furthermore, the complications arising from the overwhelming presence of the 
$c\bar c$-resonances in the $\mu\mu$-spectrum makes this measurement even more challenging. A common practice in experiment (and also in theory) is not to 
measure around the narrow resonances, $q^2\approx m_{J/\psi,\psi(2S)}^2$, while in the region of large $q^2$'s -- where the $c\bar c$-resonances are broader -- the spectrum is 
measured but for the comparison with theory it is important to integrate over a sufficiently large bin in order to rely on the quark-hadron duality. 
Contrary to that practice, as we argue in this work, it turns out that measuring $\acp(q^2)$ around the resonance region can be more beneficial than measuring it away from resonances.

In the LHCb analysis~\cite{Aaij:2016cbx} the CP-averaged $B \to K \mu^+ \mu^-$ spectrum was carefully measured including the regions close to the $c\bar c$-resonances. The spectrum was modeled by amplitude which contained Breit-Wigner parameterization of each of the resonances and the fit results revealed the strength of the resonances and the corresponding strong phases, which are a crucial ingredient to predicting the direct CP asymmetry. We use the LHCb results to interpret existing measurements of $\acp$ in the scope of BSM model with $V-A$ interaction ($\delta C_9 = -\delta C_{10}$), and in the vector current BSM model ($\delta C_9$). In both cases the effective BSM couplings are allowed to be complex in order to entail a non-zero $\acp$. The goal of this paper is to present the constraints on the complex BSM couplings from existing measurements and to show that, based on the current knowledge of resonant strong phases, $\acp (q^2)$ is enhanced in the regions close to resonances.

\section{Constraints on complex $\delta C_9$}
In this section we present how the measurements of theoretically and experimentally well understood quantities, $R_{K^{(*)}}$, $\mc{B}(B_s \to \mu^+ \mu^-)$, and CP-averaged $d\Gamma/dq^2(B \to K \mu^+ \mu^-)$, translate into allowed regions in the complex $\delta C_9$ plane. The mentioned constraints are CP-even and are only sensitive to $(\im\,\delta C_9)^2$. Furthermore we will include the constraint stemming from the measured $\acp$ of $B \to K \mu^+ \mu^-$, which depends linearly on $\im (\delta C_9)$.
Constraints on $\delta C_9$ will be presented for a BSM scenario that admits $\delta C_9 = -\delta C_{10}$, as well as for the scenario where BSM is present only in $\delta C_9$.

\subsection{$B\to K \ell^+ \ell^-$ effective Hamiltonian}
Here we remind the reader of the basic ingredients needed to compute the differential decay rate of $B\to K \ell^+ \ell^-$, where $\ell$ stands for one of the lepton flavors. As usual, the starting point is the effective Hamiltonian describing the $b \to s \ell^+ \ell^-$ transitions, namely
\begin{equation}
  \label{eq:EffLag}
\mathcal{H}^{b\to s \ell \ell}_{\mathrm{eff}} = -\dfrac{4 G_F  V_{tb}V_{ts}^\ast}{\sqrt{2}}  \sum_{i=7,9,10} C_i(\mu)\mathcal{O}_i(\mu)\,,
\end{equation}
where the short distance physics is encoded in the Wilson coefficients $C_{7,9,10}$, while the long distance part is described by the hadronic matrix elements of the effective operators
  \begin{align}
    \label{eq:O7910}
\mc{O}_7  &=  \dfrac{e m_b}{4\pi} \, (\bar{s}_R \sigma_{\mu\nu}  b_R ) F^{\mu\nu}\,,\\
  \mathcal{O}_{9(10)} &= \dfrac{e^2}{(4\pi)^2} \,(\bar{s}_L\gamma_\mu  b_L ) (\bar\ell\gamma^\mu (\gamma^5)\ell )\,.
  \end{align}
Notice that we focus here only to the operators the Wilson coefficients of which are non-zero in the Standard Model. While discussing the effects of New Physics 
relevant to $B\to K\mu^+\mu^-$ we will assume them to either modify only $C_9$ as $C_9=C_9^\mathrm{SM}+\delta C_9$, or to modify both $C_9$ and $C_{10}$ in such a way that $\delta C_9 = - \delta C_{10}$. 
The differential decay rate of $\bar B(p)\to \bar K(k)\ell^+ \ell^-$ can then be compactly written as~\cite{Bobeth:2008ij}
\begin{align}
  \label{eq:spectrum}
  {d\bar\Gamma\over dq^2} &= 2 \mc{N}(q^2)\,\biggl[ \frac{1}{6}\left( 1+ \frac{2m_\ell^2}{q^2}\right) \lambda(q^2)   \left( \vert F_V\vert^2 + \vert F_A\vert^2 \right) \cr
                      &\quad \quad\qquad+  4m_\ell^2 m_B^2  \vert F_A\vert^2 - q^2   \vert F_P\vert^2 \cr
  &\quad \quad \qquad + 2 m_\ell (m_B^2 -m_K^2-q^2) \re\left( F_P\,F_A^\ast\right) \biggr]
\,,
\end{align}
where $q^2=(p-k)^2$, $\lambda(q^2) = [q^2 - (m_B-m_K)^2] [q^2 - (m_B+m_K)^2]$, while the explicit expressions of the $q^2$-dependent functions $F_{V,A,P}$ read:
\begin{align}
F_V &= \ C_9\,  f_+(q^2) + \frac{2 m_b}{m_B+m_K} C_7\,  f_T(q^2)\,, \nonumber\\ 
F_A &= \ C_{10}\, f_+(q^2) \,,\\
F_P &= \ C_{10}\, m_\ell \left[  f_+(q^2) - \frac{m_B^2 - m_K^2}{q^2} \left(  f_0(q^2) -  f_+(q^2) \right)\right]\,. \nonumber
\end{align}
The normalization is also $q^2$-dependent:
\begin{align}
  \mc{N}(q^2) &=  \frac{G_F^2 \alpha^2 \lvert V_{tb} V_{ts}^{\ast} \rvert^2}{512 \pi^5 m_B^3}  \,  \sqrt{\lambda(q^2)  }\, \sqrt{1-\frac{4 m_\ell^2}{q^2}} \,.
\end{align}
In the above expressions, besides the Wilson coefficients $C_{7,9,10}$, we used the hadronic form factors $f_{+,0,T}(q^2)$ which parametrize the hadronic matrix elements as follows:
\begin{align}
\label{matrix-elements}
\langle K(k)| \bar{s} \gamma_{\mu} b |B(p)\rangle & =  \left[ (p + k)_\mu - {m_B^2 - m_K^2 \over q^2} q_\mu \right] f_{+}(q^2)   \,  \nonumber \\
 &  \quad  + {m_B^2-m_K^2 \over q^2} q_{\mu} f_{0}(q^2) \, ,  \\
\langle K(k)| \bar{s}\sigma_{\mu\nu}b |B(p) \rangle &= -i \left( p_\mu k_\nu  - p_\nu k_\mu \right) \frac{2 f_T(q^2)}{m_B + m_K} \, .\nonumber
\end{align}
The above form factors have been computed by means of numerical simulations of QCD on the lattice~\cite{Aoki:2019cca,Bailey:2015dka,Bouchard:2013eph} and 
we will use those results in our phenomenological discussion.

\subsection{Resonant $B \to K \ell^+ \ell^-$ spectrum}

A non-resonant contribution of the $c\bar c$-pairs, as well as those arising from the light quarks, is usually included by promoting the Wilson coefficient $C_9$ to $C_9(q^2)= C_9+ Y(q^2)$ where, owing to the quark-hadron duality, the function $Y(q^2)$ is computed perturbatively. That obviously cannot account for the $c\bar c$-resonances, present in the $q^2$-spectrum % =(p_{\ell^+}+p_{\ell^-})^2$
of the decay. For that reason, when comparing theory with experiment, one vetoes regions around the prominent resonances, such as $J/\psi$, $\psi(2S)$ and $\psi(3770)$, or by working in the low $q^2$ region ($q^2 \lesssim m_{J/\psi}^2$) in order to avoid the $c\bar c$-resonances altogether.   
Since the purpose of this work is to discuss the potential effects of CP-violation, we will not follow the usual description but, instead, we will adopt the model of Ref.~\cite{Aaij:2016cbx} in which the authors actually reconstructed contributions from the $c \bar c$-resonances through the fit to the experimental data.~\footnote{In addition to the $c\bar c$-resonances they were able to resolve the contributions coming from the light quark resonances, such as $\rho$, $\omega$, $\phi$, cf. Ref.~\cite{Aaij:2016cbx} for more details.} More specifically they trade $C_9$ for
\begin{align}
  \label{eq:1}
  C_9^\mrm{eff} (q^2) &= C_9 + C_9^\mathrm{res}(q^2)  \nonumber\\
  &= C_9 +  \sum_{j} \frac{m_j \,\Gamma_j \,\eta_j \, e^{i \delj}}{m_j^2-q^2 - i\, m_j \Gamma_j(q^2)},
\end{align}
with $j \in\{J/\psi, \psi(2S),\psi(3770),\psi(4040), \psi(4160), \psi(4415)\}$, for which the masses ($m_j$) and widths ($\Gamma_j$) are well known~\cite{Zyla:2020zbs}.
The $q^2$-dependent width function reads $\Gamma_j(q^2) = \Gamma_j \times \sqrt{1-4m_\ell^2/q^2}/ \sqrt{1-4m_\ell^2/m_i^2}$.
By using $C_9^\mrm{eff} (q^2)$ of Eq.~\eqref{eq:1} to fit the measured CP-averaged spectrum the authors of Ref.~\cite{Aaij:2016cbx} were able to determine the value of $\eta_j$ 
and the strong phase $\delta_j$ for each of the six $c\bar c$-resonances. The Wilson coefficients $C_9$ and $C_{10}$, assumed to be real, have been also fitted in the analysis, with their values in line with results of the global fits. It appears though that there is a fourfold ambiguity related to the choice of the signs of the first two resonances. By using the results for all $\eta_j$ and $\delj$ from Ref.~\cite{Aaij:2016cbx} 
we were able to reconstruct their model. Notice also that in Eq.~\eqref{eq:1}, on the right hand side, we have $C_9=C_9^\mathrm{SM}+\delta C_9$. While the authors 
of Ref.~\cite{Aaij:2016cbx} assumed $\delta C_9$ to be real, we will allow it to be complex. In other words, we allow for a possible BSM weak phase.

\subsection{$R_{K^{(*)}}$ and $\mc{B}(B_s \to \mu^+ \mu^-)$}

%%%%%%%%%%%%%%%%%%%%%%%%%%%%%%%%%%%%%%%%%%%%%%%%%
%%%%%%%%%%%%%%%%%%%%%%%%%%%%%%%%%%%%%%%%%%%%%%%%%
\begin{figure}[!h]
  \centering 
  \vspace{0.1cm}
\includegraphics[scale=0.55]{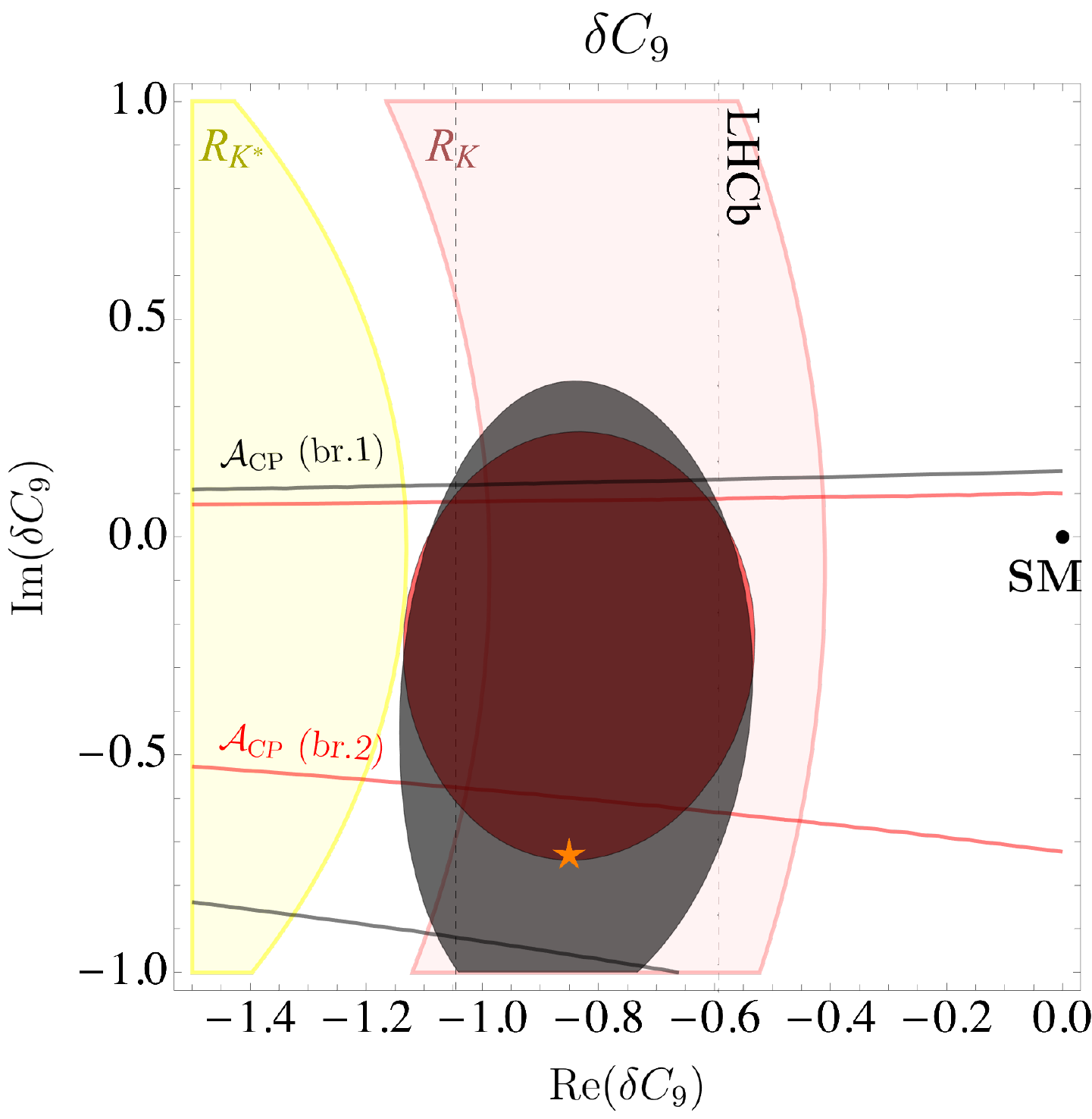}
\caption{Constraints on $\bigl( \re(\delta C_9), \im(\delta C_9)\bigr)$ compatible with $R_K^\mathrm{exp}$ and $R_{K^*}^\mathrm{exp}$, as well as with the error bars on $\acp$ measured in the bins between $2 \e{GeV}^2 < q^2 < 8\e{GeV}^2$, as reported in Ref.~\cite{Aaij:2014bsa}. The horizontal lines correspond to the $\acp$ constraint for branches 1 and 2, see text for details. Vertical dashed lines enclose the $2\sigma$ region of $\re(\delta C_9)$ obtained from CP-averaged spectra in Ref.~\cite{Aaij:2016cbx}.
  The scenario considered in this case is the one in which all of the BSM effects are described by $\delta C_9$ only. The fitted $1\sigma$ values of $\delta C_9$ are ellipses in gray~(red) for strong phases in branch 1~(2).  By a star we denote our benchmark point which we chose to be $\delta C_9 = -0.85-0.73 i$.}
  \label{fig:C9}
\end{figure}
%%%%%%%%%%%%%%%%%%%%%%%%%%%%%%%%%%%%%%%%%%%%%%%%%
%%%%%%%%%%%%%%%%%%%%%%%%%%%%%%%%%%%%%%%%%%%%%%%%%
As we already mentioned above, we consider two scenarios of New Physics. In the first one we will allow only $C_9$ to receive an extra contribution, $\delta C_9  \in \mathbb{C}$, while leaving 
other Wilson coefficients at their Standard Model values. In such a situation it suffices to use $R_{K^{(\ast )}}$ to determine $\delta C_9$. Knowing that $R_K^\mathrm{exp}$ and $R_{K^*}^\mathrm{exp}$ are obtained from the partial branching fractions integrated in the interval $q^2\in [1.1,6]~\gev^2$, we obtain the simple formulas: 
  \begin{subequations}
      \label{eq:RKC9}
    \begin{align}
      R_K &= 1.003 + 0.244\, \re(\delta C_9)\nn\\
      &\quad+ 4.01\E{-3}\, \im(\delta C_9)
        + 0.028\, |\delta C_9|^2,\\
      R_{K^*} &= 0.997 + 0.202\,\re(\delta C_9)\nn\\
       &\quad+  1.65\E{-3}\,  \im(\delta C_9)
      + 0.033\, |\delta C_9|^2, 
    \end{align}
  \end{subequations}
in both of which most of the $1\%$ overall error affects the first term. Using these two expressions, together with $R_K^\mathrm{exp}$ and $R_{K^*}^\mathrm{exp}$, we get a region of allowed values for $\bigl( \re(\delta C_9), \im(\delta C_9)\bigr)$ shown in Fig.~\ref{fig:C9}.

%%%%%%%%%%%%%%%%%%%%%%%%%%%%%%%%%%%%%%%%%%%%%%%%%
%%%%%%%%%%%%%%%%%%%%%%%%%%%%%%%%%%%%%%%%%%%%%%%%%
\begin{figure}[!h]
  \centering 
  \vspace{0.1cm}
\includegraphics[scale=0.55]{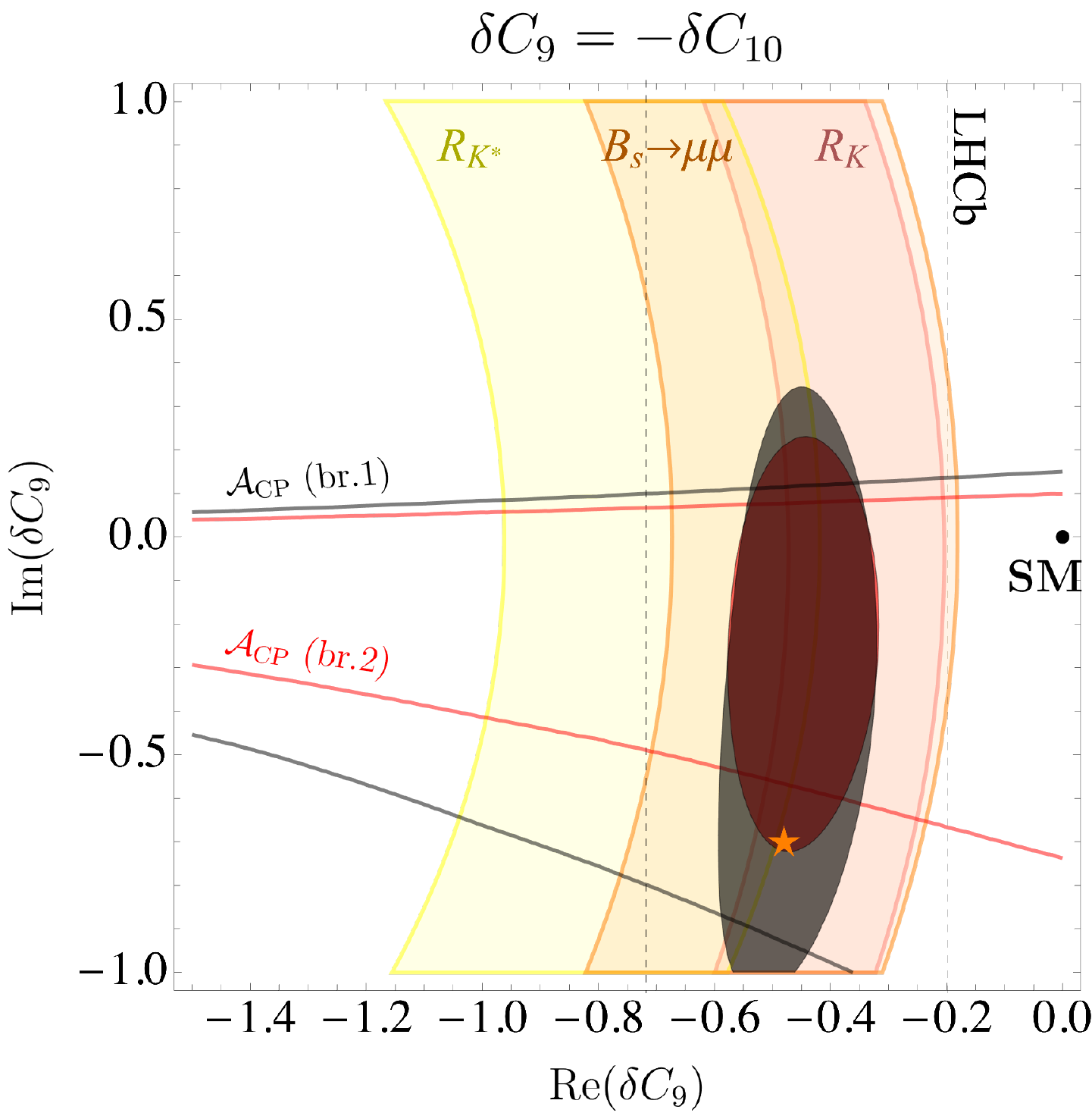}
 \caption{In addition to the constraints mentioned in the caption of Fig.~\ref{fig:C9}, here we also account for $\mathcal{B}(B_s\to\mu^+\mu^-)^\mathrm{exp}$. The scenario we consider in this plot is the one where the BSM effects are described by $\delta C_9$ and $\delta C_{10}$, satisfying $\delta C_9=-\delta C_{10}$. Our benchmark point in this case, $\delta C_9 = -0.48-0.7 i$, is depicted by a star.}
  \label{fig:C910}
\end{figure}
%%%%%%%%%%%%%%%%%%%%%%%%%%%%%%%%%%%%%%%%%%%%%%%%%
%%%%%%%%%%%%%%%%%%%%%%%%%%%%%%%%%%%%%%%%%%%%%%%%%
In the second scenario considered in this work, we allow both $C_{9,10}$ to receive contributions from BSM, $C_{9,10}=C_{9,10}^\mathrm{SM}+\delta C_{9,10}$, but by respecting the 
left-handedness, i.e. $\delta C_9 = -\delta C_{10}$. In this case, the above formulas for $R_{K^{(*)}}$ become:
  \begin{subequations}
      \label{eq:RKC9C10}
    \begin{align}
      R_K &= 1.003 + 0.477\, \re(\delta C_9)\nn\\
      &\quad+ 4.01\E{-3}\, \im(\delta C_9) 
        + 0.057 |\delta C_9|^2,\\
      R_{K^*} &= 0.997 + 0.472 \,\re(\delta C_9)\nn\\
       &\quad+ 1.65\E{-3}\, \im(\delta C_9)
      + 0.066 |\delta C_9|^2. 
    \end{align}
  \end{subequations}
Since we allow in this scenario $\delta C_{10}\neq 0$, the recently updated $\mathcal{B}(B_s\to\mu^+\mu^-)^\mathrm{exp}=(2.69^{+0.37}_{-0.35})  \E{-9}$~\cite{LHCb-CONF-2020-002} becomes an important constraint too. Notice that $\mathcal{B}(B_s\to\mu^+\mu^-)^\mathrm{exp}$ is $2.5\,\sigma$ smaller than predicted in the SM, $\mathcal{B}(B_s\to\mu^+\mu^-)^\mathrm{SM}=(3.66\pm 0.11)  \E{-9}$~\cite{Beneke:2019slt}. For the purpose of determining $\delta C_{10} \in \mathbb{C}$, we use 
\begin{align}
\label{eq:BS}
\mathcal{B}(B_s\to \mu^+\mu^-)^\mathrm{th} &= \tau_{B_s}\dfrac{\alpha^2 G_F^2 m_{B_s}}{16 \pi^3}  \left| V_{tb}V_{ts}^\ast \right|^2  m_\mu^2   \nonumber \\
& \quad\times \sqrt{1-\frac{4 m_\mu^2}{m_{B_s}^2}} \left|C_{10}\right|^2 f_{B_s}^2,
\end{align}
where, thanks to the lattice QCD efforts, the uncertainty in the decay constant is not anymore an obstacle to constraining the BSM contribution, $f_{B_s}= (230.3\pm 1.3)\,\mev$~\cite{Aoki:2019cca}. 
In order to compare Eq.~\eqref{eq:BS} to $\mathcal{B}(B_s\to\mu^+\mu^-)^\mathrm{exp}$, one also needs to account for the effect of $B_s- \overline{B}_s$ oscillations which, to a good approximation, amounts to~\cite{DeBruyn:2012wj}
\begin{equation}
\mathcal{B}(B_s\to\mu^+\mu^-)^\mathrm{exp} \approx \dfrac{1}{1-y_s}\mathcal{B}(B_s\to \mu^+\mu^-)^\mathrm{th},
\end{equation}
\noindent where $y_s=\Delta \Gamma_{B_s}/(2 \Gamma_{B_s})=0.061(7)$~\cite{Aaij:2014zsa}. 
Finally, a comparison between theory and experiment in the scenario with $\delta C_9 = -\delta C_{10}$ results in a region of allowed values in the $\bigl( \re(\delta C_9), \im(\delta C_9)\bigr)$ plane, which we plot in Fig.~\ref{fig:C910}. We should also note that a possibility of the complex Wilson coefficients was recently discussed in Refs.~\cite{Biswas:2020uaq,Calibbi:2019lvs}.

\subsection{$\acp$ constraint}
The measurement of $\acp$ in non-resonant regions of $B^\pm \to K^\pm \mu^+ \mu^-$ has been presented by the LHCb collaboration in~\cite{Aaij:2014bsa}. 
We will employ their bin-by-bin results for $\acp$ in the region of $2 \e{GeV}^2 < q^2 < 8\e{GeV}^2$, with each bin-width being $1\e{GeV}^2$. With binned data the CP-asymmetry is defined as
\begin{align}
  \acp(\mrm{bin}) &= \frac{\bar\Gamma(\mrm{bin}) - \Gamma(\mrm{bin})}{\bar\Gamma(\mrm{bin}) + \Gamma(\mrm{bin})}\,,\\
  \mrm{bin} &\equiv [q_1^2,q_2^2]\,\nn,
\end{align}
where $\bar\Gamma\, (\Gamma)$ refers to decay
$B^- \to K^- \mu^+ \mu^-\,(B^+ \to K^+ \mu^+ \mu^-)$. Importantly, the
theoretical prediction of $\acp$ in each of the
$[2,3]\e{GeV}^2,\ldots,[7,8]\e{GeV}^2$ bins depends on the resonant
spectrum too and therefore the results of Ref.~\cite{Aaij:2016cbx}
should be accounted for carefully.  More specifically, as we shall see
below, the theoretical prediction for $\acp$ is proportional to
$\im (\delta C_9)\times \im (C_9^\mrm{res})$ in a given bin.  With the
fourfold sign ambiguity regarding the first two resonances, as
reported in Ref.~\cite{Aaij:2016cbx}, it turns out that it suffices to
consider two distinct solutions with negative strong phase
$\delta_{J/\psi}$ and either negative ({\sl ``Branch 1''}) or positive
value ({\sl ``Branch 2''}) of the strong phase $\delta_{\psi(2S)}$:
\begin{subequations}
  \label{eq:branches}
  \begin{align}
    \textrm{Branch 1:  }\quad \delta_{J/\psi} &= -1.66\,,\quad \delta_{\psi(2S)} = -1.93\,,\\
    \textrm{Branch 2:  }\quad \delta_{J/\psi} &= -1.50\,,\quad \delta_{\psi(2S)} = 2.08\,.
  \end{align}  
\end{subequations}
The
two fit branches are shown in Figs.~\ref{fig:C9},\ref{fig:C910}.  The
other two solutions would result in flipping the sign of
$\im (\delta C_9)$. The allowed region of $\delta C_9$, obtained from
the fit to binned $\acp$ in the region $2\,\gev^2 \leq q^2\leq 8 \gev^2$, results in
horizontal lines in Figs.~\ref{fig:C9} and \ref{fig:C910}.\footnote{We
  have treated the experimental values of $\acp$ in different bins as uncorrelated.}

\section{Behavior of direct CP asymmetry at the resonance}
The differential CP asymmetry is defined as
\begin{equation}
  \label{eq:acp}
  \acp(q^2) = \frac{d\bar\Gamma/dq^2-d\Gamma/dq^2}{d\bar\Gamma/dq^2+d\Gamma/dq^2},
\end{equation}
where, as mentioned above, $d\bar \Gamma/dq^2$ and $d\Gamma/dq^2$ refer to the differential decay rates of $\bar B \to \bar K \mu^+ \mu^-$ and $B \to K \mu^+ \mu^-$. In the following, for simplicity, we will neglect the muon mass in 
Eq.~\eqref{eq:spectrum} and set $C_{10} = C_{10}^\mrm{SM} + \delta C_{10}$, $C_9 = C_9^\mrm{SM} + \delta C_9$,
$C_7 = C_7^\mrm{SM}$, to get:
\begin{equation}
  \label{eq:RateSimply}
  \begin{split}
    \frac{d\bar\Gamma}{dq^2} &= \frac{\mc{N} \lambda}{3} \left[ f_+(q^2) \right]^2 \Bigg\{  |C_{10}^\mrm{SM} + \delta C_{10}|^2 \\
    + &\left| C_9 + \cres + \frac{2m_b}{m_B + m_K} \frac{f_T(q^2)}{f_+(q^2)} C_7^\mrm{SM} \right|^2 \Bigg\},
  \end{split}
\end{equation}
where, again for notational simplicity, we omit the argument in $\mc{N}$ and $\lambda$. Note that $C_9^\mrm{res}(q^2)$ is complex and contains a CP-even (strong) phase, cf. Eq.~\eqref{eq:1}. $C_{10}^\mrm{SM}$ and $C_7^\mrm{SM}$ are assumed to be real so that the only potential sources of CP violation are in the imaginary parts of $\delta C_{9,10}$. The rate for CP-conjugated decay is obtained from~\eqref{eq:RateSimply} by replacing $\delta C_{9,10} \to \delta C_{9,10}^\ast$. The necessary ingredient for a non-zero $\acp$ is interference between two terms of the amplitude which have different strong and weak phases. Such effect is possible only for $\im(\delta C_9)$ that interferes with $\im (\cres)$ and drives the numerator of $\acp$, namely:
\begin{equation}
  \label{eq:aCPnumerator}
  \frac{d\bar \Gamma}{dq^2} -  \frac{d\Gamma}{dq^2} = \frac{4\mc{N} \lambda}{3}\left[f_+(q^2)\right]^2\, \im (\cres) \,\im (\delta C_9)\,.
\end{equation}
%%%%%%%%%%%%%%%%%%%%%%%%%%%%%%%%%%%%%%%%%%%%%%%%%
%%%%%%%%%%%%%%%%%%%%%%%%%%%%%%%%%%%%%%%%%%%%%%%%%
\begin{figure}[!t]
  \centering
  \vspace{0.1cm}
  \includegraphics[scale=0.55]{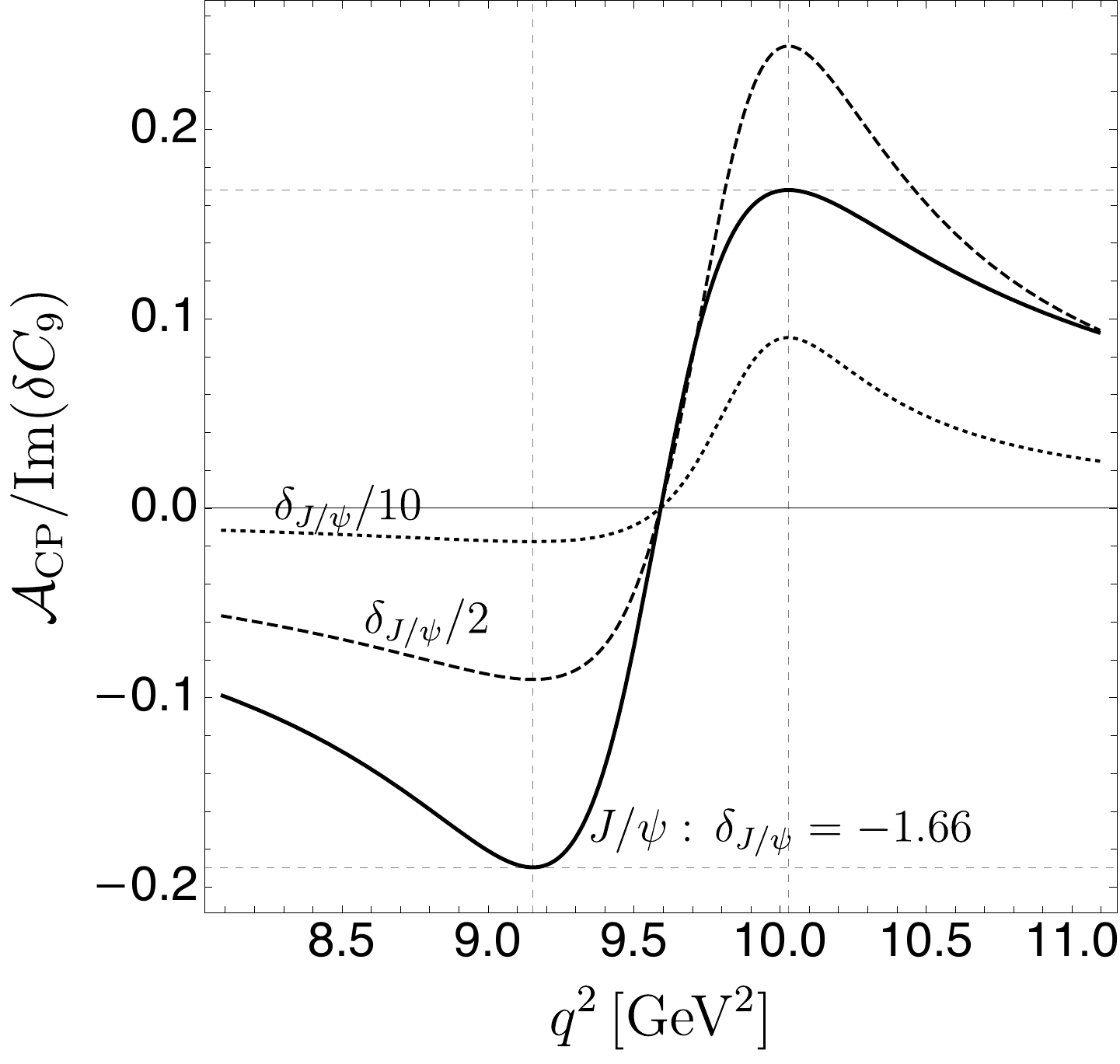}
  \caption{Behavior of $\acp\equiv\acp(q^2)$ around the $J/\psi$-resonance for which $\eta_{J/\psi}=8500$, $\delta_{J/\psi} = -1.66$. Due to the large strong phase the $\acp$ is antisymmetric with respect to the position of the $J/\psi$ peak, $q^2=9.58~\gev^2$. Its maximal values are attained far away from the peak (full line). We also show the behavior of $\acp$ if the strong phase was 2- (dashed) or 10-times (dotted) smaller. Dashed gridlines denote the positions and heights of $\acp$ extrema according to Eqs.~\eqref{eq:extrema}.}
  \label{fig:acpToy}
\end{figure}
%%%%%%%%%%%%%%%%%%%%%%%%%%%%%%%%%%%%%%%%%%%%%%%%%
%%%%%%%%%%%%%%%%%%%%%%%%%%%%%%%%%%%%%%%%%%%%%%%%%
The contribution of $\delta C_{10}$ is not important, as it only modifies the denominator of $\acp$ and we neglect it in this discussion. To make the argument clearer, let us now assume that $\delta C_9$ is strictly imaginary and that $|\delta C_9|^2$ is negligible with respect to $|C_9^\mrm{SM}|^2$ in the denominator of $\acp(q^2)$. We are interested in the behavior of $\acp\equiv \acp(q^2)$ close to one of the aforementioned $\bar c c$-resonances, where $\cres$ is approximately
\begin{equation}
\cres \approx \frac{m_j \Gamma_j \eta_j e^{i\delj}}{m_j^2 - q^2 -i m_j \Gamma_j}\,.
\end{equation}
The expression for $\acp$ then boils down to
\begin{equation}
  \label{eq:aCP}
  \begin{split}
    \acp &= \im (\delta C_9)\,  \frac{2\eta_j \left(\cos \delj - x \sin \delj\right)}{\eta_j^2 - 2 \eta_j B \left[\sin \delj + x  \cos \delj \right] + A \left[1+ x^2 \right] }\,,
  \end{split}
\end{equation}
where $x \equiv (q^2-m_j^2)/(m_j\Gamma_j)$ measures the distance from the resonance peak, while $B$ and $A$ are defined as
\begin{align}
  B &= C_{9}^\mrm{SM} + \frac{2m_b}{m_B+m_K} \frac{  f_T(q^2)}{ f_+(q^2) }  C_7^\mrm{SM} \approx 3.8\,,\nn\\
  A &= (C_{10}^\mrm{SM})^2+B^2 \approx 31\,.
\end{align}
The above two quantities are almost constant throughout the whole range of physical $q^2$'s.
If the strong phase is large, $|\delta_j| \approx \pi/2$, then the imaginary part of $\cres$ vanishes on the resonant peak, leading to zero $\acp$. In the limit of small strong phase, $\delj = 0$, the value of the asymmetry on the peak is
\begin{equation}
  \label{eq:aCPpeak}
\bigl. \acp (x \to 0) \biggr|_{\delj = 0} = \im (\delta C_9)\, \frac{2 \eta_j}{\eta_j^2 + A}.
\end{equation}
The lowest lying and the most prominent $c\bar c$-resonances are $J/\psi$ and $\psi(2S)$. Their narrow widths and relatively large branching 
fractions to the $\mu^+ \mu^-$ channel imply large $\eta_j$ parameters [$\eta_{J/\psi} \approx 8.5\E{3}$, $\eta_{\psi(2S)} \approx 1.4\E{3}$].
Therefore, $\acp$ is suppressed by $\eta_j$ at the resonant peak. Farther away from the peak, however, $\acp$ is enhanced by the large values of $\eta_j$, as can be seen in the following limit
\begin{equation}
  \label{eq:aCPaway}
\acp (|x| \to \infty) = \im (\delta C_9)\, \frac{2 \eta_j (\cos \delj -x \sin \delj)}{ A \, x^2}\,.
\end{equation}
In the large $x$ regime the leading asymptotic term is proportional to $\sin \delj$,
\begin{align}
  \label{eq:acpasympt}
  \acp &\approx \im (\delta C_9)\,\frac{-2\sin \delj \,\eta_j}{Ax} \cr 
  &=  \im (\delta C_9) \, \frac{-2\sin \delj\,\eta_j \Gamma_j m_j}{A(q^2-m_j^2)}\,.
\end{align}
Finally, $\acp$ also develops two extrema around the resonant peak. Explicit expression for their positions ($q^2_{1,2}$) can be derived from Eq.~\eqref{eq:aCP}. We obtain:
\begin{subequations}
  \label{eq:extrema}
  \begin{align}
    \label{eq:acppos}
    q^2_{1,2} &= m_j^2  \pm m_j \Gamma_j \left(\frac{\eta_j}{\sqrt{A}} +\frac{B}{\sqrt{A} \sin \delj}\right)\nn \\
              &\quad\qquad + m_j \Gamma_j \cot \delj + \mc{O}(1/\eta_j)\,, \\
    \acp \left(q^2_{1,2}\right) &= \im ( \delta C_9 )\,\frac{\sin \delj}{\sqrt{A} \pm B \cos\delj}+\mc{O}(1/\eta_j)\,.                 \label{eq:acpmax}
  \end{align}
\end{subequations}
To leading order in $1/\eta_j$ the positions of the extrema of $\acp$ depend on the product $\eta_j \Gamma_j$. The maximum value of $\acp$, instead, depends only on the strong phase, as can be seen from Eq.~\eqref{eq:acpmax}. The conclusion that we can draw from this analysis is that the narrow resonances with large $\eta_j$ 
enhance $\acp$ much farther away from the resonant peak than one would na\"ively expect from the small resonance width. This feature is explicitly shown in Fig.~\ref{fig:acpToy} where $\acp$ is plotted around the $J/\psi$-resonance. As we see the extrema of $\acp$ are positioned at $x \approx \pm 1500$, i.e. at $q^2 - m_{J/\psi}^2 \approx \pm 1500\ m_{J/\psi} \Gamma_{J/\psi}$.
A second message is that a large strong phase $\delj$ makes the distribution of $\acp$ antisymmetric around the peak, whereas with smaller
strong phase values the shape of $\acp$ around the peak would become asymmetric and its size much smaller. Qualitative features of $\acp$ around the $\psi(2S)$ are similar to those presented in Fig.~\ref{fig:acpToy} for $J/\psi$, as both these resonances have large $\delj$, barring a sign ambiguity, as determined in Ref.~\cite{Aaij:2016cbx}. Thus, in the presence of the imaginary part of $\delta C_9$ the $\acp$ would be enhanced around these two resonances. Furthermore, the shape of $\acp (q^2)$ would be approximately antisymmetric with respect to each of the resonant peaks.

\begin{figure*}[!htb]
  \centering
  \vspace{0.1cm}
  \includegraphics[scale=0.55]{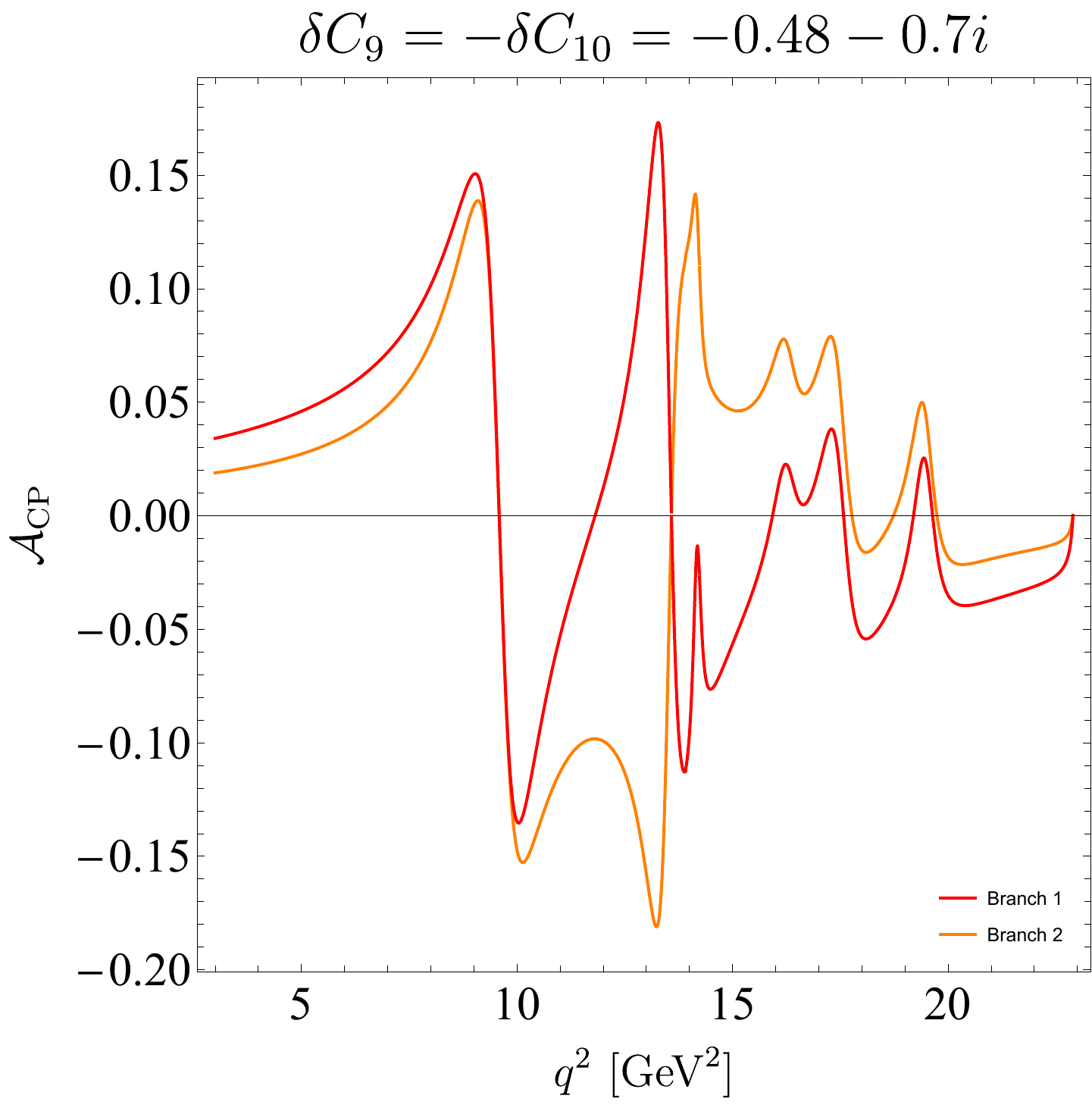}
  \includegraphics[scale=0.61]{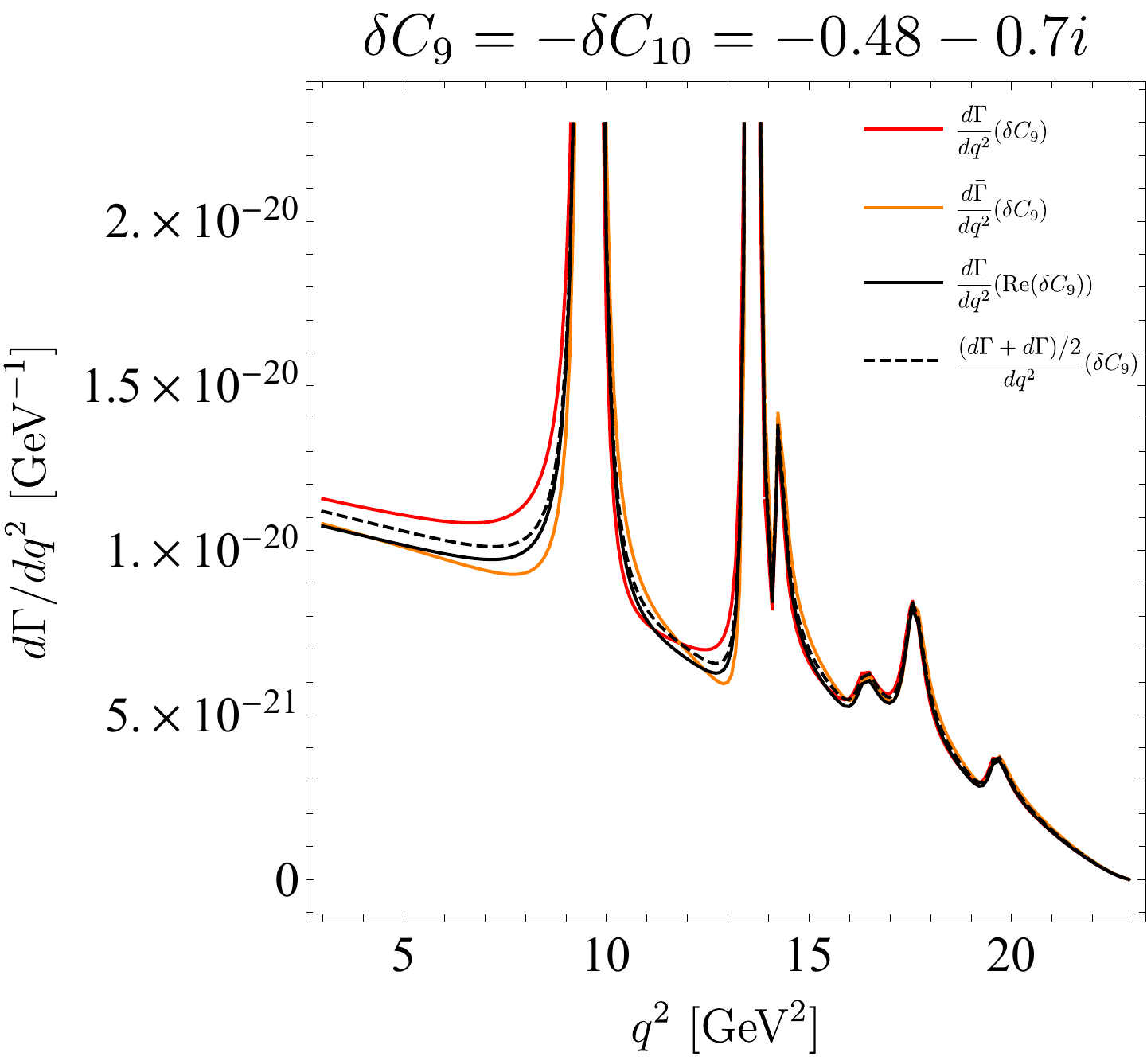}
  \caption{In the left panel we plot $\acp\equiv \acp(q^2)$ in the full physical region, and for the benchmark point in the BSM scenario satisfying $\delta C_9 = - \delta C_{10}$.  
Two branches correspond to two sets of solutions for strong phases obtained from the fit to the data of Ref.~\cite{Aaij:2016cbx}. Other two branches would correspond to flipping the sign of the above-depicted $\acp(q^2)$ in the region $q^2\lesssim m_{\psi(2S)}^2$. In the right panel we plot the CP-averaged differential decay rate in this scenario to show that the effect of $\delta C_9\in \mathbb{C}$ is extremely difficult to disentangle from this quantity. Indeed, the (dashed) curve corresponding to the CP-average of decay widths and $\im\left(\delta C_9\right) = - 0.7$ is hardly distinguishable from the full black curve in which $\im\left(\delta C_9\right)$ is set to zero. Here the {\sl``branch 1''} set of resonant parameters has been used.}
  \label{fig:acpfull}
\end{figure*}

\subsection{Shape and size of $\acp (q^2)$ for benchmark values of $\delta C_9$}

So far in this Section we introduced several simplifications to make our discussion clearer. We now use the full formulas to draw $\acp (q^2)$ for the two benchmark points we have chosen in both of our scenarios, denoted by stars in Figs.~\ref{fig:C9},\ref{fig:C910}. These benchmark points correspond to a large value of $|\im( \delta C_9)|$. 
Following the line of the discussion above, besides a non-zero value of $\im( \delta C_9)$, it is essential to have a good handle over $\im(\cres)$, which we were able to get thanks to the results of Ref.~\cite{Aaij:2016cbx} in which 
the resonant parameters are given for four possible solutions (branches). In the region $q^2 \lesssim m_{\psi(2S)}^2$ only two of those would give different shapes of $\acp (q^2)$, while the other two solutions would simply flip the overall sign of those $\acp (q^2)$ that we already obtained from the first two branches. Regarding the first two branches the results are plotted in Fig.~\ref{fig:acpfull}. We note immediately that due to the fact that $\acp (q^2)$ essentially depends on $\im( \delta C_9)$, and since in both benchmark points $\im( \delta C_9)\approx -0.7$, the two scenarios for each branch practically coincide. This is why we decided to plot $\acp (q^2)$ only in the scenario with $\delta C_9 = - \delta C_{10}$. 
We need to emphasize once again that it is essential to measure $\acp$ in a bin before and in another bin after the peak. Another remark is that locally, near the first two resonances, the value of $\acp (q^2)$ can be appreciable, about $\pm 15\%$ for our benchmark $\im( \delta C_9)$.  
It should be reiterated, however, that the effects of CP-violation would be extremely difficult to disentangle from the differential CP-averaged decay width, as we show also in Fig.~\ref{fig:acpfull}.

\section{Conclusion}
In most of the phenomenological studies of the exclusive $B\to K^{(\ast )}\ell^+\ell^-$ decays, the potential effects of New Physics are described by the shift of some of the Wilson coefficients. 
From the experimental data on $B\to K^{(\ast )}\mu^+\mu^-$ the most favored scenarios seem to be those modifying $C_9$, or those in which both $C_9$ and $C_{10}$ are modified but in such a way that $\delta C_9 = - \delta C_{10}$, where $\delta C_{i}$ refers to the BSM contribution. So far this shift was considered to be real-valued in order to make the number of parameters minimal. 
However, the BSM contributions can be complex, $\delta C_{i} \in \mathbb{C}$, and the measurement of the CP-asymmetry could reveal the presence of that BSM (weak) phase.  
From the current data regarding the $B$-anomalies, namely $R_{K^{(\ast)}}^\mathrm{exp} < R_{K^{(\ast)}}^\mathrm{SM}$, as well as from the measured $\mathcal{B}( B_s\to \mu^+\mu^-)$, 
we were able to constrain the regions in the $\bigl(\re(\delta C_9), \im(\delta C_9)\bigr)$ plane, clearly showing that having $\im(\delta C_9)\neq 0$ is, in both scenarios, perfectly plausible and consistent with data. 
That effect, however, is small and its measurement along the lines presented in Ref.~\cite{Aaij:2014bsa} would require extremely high experimental precision. 
In this work we showed that, if measured closely to the peaks of the $c\bar c$-resonances, the effects of $\im(\delta C_9)\neq 0$ can be further amplified by the strong phase of 
each resonance and the resulting CP-asymmetry, $\acp(q^2)$, measured before and/or after the resonance's peak can be easier to distinguish. 
We showed, through a simplified example, the details of how this enhancement actually occurs. 
We focused on the $B\to K \mu^+\mu^-$ decay, but the discussion can straightforwardly be extended to $B\to K^\ast \mu^+\mu^-$.~\footnote{In fact, after completing this work, we became aware that the research along these lines for the case of $B\to K^\ast \mu^+\mu^-$ has been presented in Ref.~\cite{Blake:2017fyh}.}

\section*{Acknowledgmens}
This project was partly funded by the EU's Horizon 2020 research and innovation program under the Marie Sklodowska-Curie grant agreement No.860881 (ITN HIDDeN). The project was in part financially supported by the Slovenian Research Agency (research core funding No. P1-0035 and J1-8137). This article is based upon work from COST Action CA16201 PARTICLEFACE supported by COST (European Cooperation in Science and Technology). A.~S. is supported by the Young Researchers Programme of the Slovenian Research Agency under the grant No. 50510, core funding grant P1-0035.
N.~K. would like to thank ``Laboratoire de Physique des 2 Infinis Ir\`ene Joliot-Curie'', Orsay, for hospitality and support during a visit which led to the completion of this work.

\bibliography{paper_CP}

\end{document}